\documentclass[a4paper,fleqn]{cas-sc}

\usepackage{lineno}
\usepackage[numbers]{natbib}
\usepackage{amsmath}
\usepackage{url}
\usepackage{graphicx}
\usepackage{siunitx}
\usepackage{color} 
\usepackage{placeins} 
\usepackage{ulem}
\usepackage{amsmath}
\usepackage{booktabs}
\usepackage{multirow}

\def\tsc#1{\csdef{#1}{\textsc{\lowercase{#1}}\xspace}}
\tsc{WGM}
\tsc{QE}
\definecolor{royalpurple}{rgb}{0.47, 0.32, 0.66}

\newcommand{\ESix}{e6-diamond}
\newcommand{\KU}{KU-diamond}
\newcommand{\Am}{$^{241}\mathrm{Am}$}
\newcommand{\Ba}{$^{133}\mathrm{Ba}$}
\newcommand{\Cd}{$^{109}\mathrm{Cd}$}

\newcommand{\keV}{\,$\mathrm{keV}$}
\newcommand{\MeV}{\,$\mathrm{MeV}$}

\begin{document}
\let\WriteBookmarks\relax
\def\floatpagepagefraction{1}
\def\textpagefraction{.001}
\shorttitle{Radiation tolerance of a diamond radiation detector for space use}

\shortauthors{Ando et al.}  

\title [mode = title]{Radiation tolerance of a diamond radiation detector for space use}



%

\author[1]{Yoshiyuki Ando} 

\corref{cor1}

 \fnmark[*]

\ead{aqingzhi@gmail.com}



\affiliation[1]{organization={College of Science and Engineering, Kanazawa University},
            addressline={Kakuma}, 
            city ={Kanazawa},
            state={Ishikawa},
            postcode={920-1192},
            country={Japan},
            }

\author[2,3]{Shutaro Ueda}[
orcid=0000-0001-6252-7922
]
\affiliation[2]{
	organization={Faculty of Mathematics and Physics, Institute of Science and Engineering, Kanazawa University},
	addressline={Kakuma}, 
	city={Kanazawa},
	state={Ishikawa}, 
	postcode={920-1192},
	country={Japan},
	}
\affiliation[3]{
	organization={Advanced Research Center for Space Science and Technology (ARC-SAT), Kanazawa University},
	addressline={Kakuma}, 
	city={Kanazawa},
	state={Ishikawa}, 
	postcode={920-1192},
	country={Japan},
	}

\affiliation[4]{organization={Institute of Space and Astronautical Science},
            addressline={3-1-1 Yoshinodai, Chuo-ku}, 
            city={Sagamihara},
            state={Kanagawa},
            postcode={252-5210},
            country={Japan}}

 \fnmark[*]

\ead{shutaro@se.kanazawa-u.ac.jp}




\affiliation[5]{organization={Advanced Research Center for Diamond Science and Technology (ARCDia),
Kanazawa University},
            addressline={Kakuma}, 
            city ={Kanazawa},
            state={Ishikawa},
            postcode={920-1192},
            country={Japan},
            }

\affiliation[6]{organization={Graduate School of Natural Science and Technology, Kanazawa University},
            addressline={Kakuma}, 
            city ={Kanazawa},
            state={Ishikawa},
            postcode={920-1192},
            country={Japan},
            }

\affiliation[7]{organization={The Wakasa Wan Energy Research Center},
            addressline={64-52-1, Nagata}, 
            city={Tsuruga},
            state={Fukui},
            postcode={914-0135},
            country={Japan},
            }

\cortext[1]{Corresponding author}

\makeatletter\def\Hy@Warning#1{}\makeatother

\author[1]{Ryota Heibatake}
\author[1]{Kaito Ozawa}
\author[2,3]{Makoto Arimoto}
\author[2,3]{Tatsuya Sawano}
\author[1,3,4]{Daisuke Yonetoku}

\author[5,6]{Kimiyoshi Ichikawa}
\author[5,6]{Norio Tokuda}
\author[6]{Taichi Miyazaki}

\author[6]{Shoya Matsuda}
\author[6]{Yasuhiro Shoji}

\author[7]{Satoshi Hatori}
\author[7]{Kyo Kume}
\author[7]{Shinko Sando}
\author[7]{Satoshi Mizushima}


\begin{abstract}
We present a study of the radiation tolerance of two types of diamond radiation detectors for space use. We plan to launch a 3U-size CubeSat, KSAT3-X, developed by Kanazawa University in 2027. The KSAT3-X mission is aimed to observe inflows and outflows of charged particles such as electrons and protons, particularly in the $10 - 40$\,keV energy range, in the Earth's magnetosphere. As the mission instrument, we have developed two diamond radiation detectors. The first is composed of a microwave plasma chemical vapor deposition (MPCVD) diamond fabricated by Element Six, and the second is based on a MPCVD diamond produced in-house at Kanazawa University. We irradiate both diamonds with 100\,MeV protons and evaluate their spectroscopic performance as an indicator of radiation tolerance using characteristic X-rays from radioisotope sources. We find no significant degradation in their spectroscopic performance up to at least the 10-year equivalent irradiation under the orbital environments of KSAT3-X. We additionally irradiate the Element Six diamond with 100\,MeV protons up to the 100-year equivalent. As a result, no significant degradation in the spectroscopic performance is observed. These results indicate that the two diamond radiation detectors have sufficiently high radiation tolerance. We also discuss possible physical origins of the observed difference in the spectroscopic performance between the two detectors.
\end{abstract}




\maketitle
\section{Introduction}\label{sec:Intro}

The Earth is an ideal laboratory not only for understanding ecosystems in its magnetosphere but also for exploring those of (exo)planets. Since the discovery of O$^{+}$ ions in the Earth's magnetosphere \cite{Shelley72}, observations of charged particles such as electrons, protons, and ions have revealed various phenomena occurring in the ecosystems \citep[see e.g.,][for reviews]{Yau1997, Yau07}.

It has been known that there are two major components of charged particles in the magnetosphere. One component is based on charged particles originating from the Sun via solar winds, while the other is composed of those escaping from the Earth's atmosphere, i.e., outflows into space. The outflows of ions from the Earth's atmosphere were first discovered by observations of ISIS-2 \cite{Hoffman74}. Since then, it has been recognized that outflows play an important role in the Earth's magnetospheric ecosystem. In fact, O$^{+}$ ions are transported to the Moon from the Earth's atmosphere \cite{Terada17}. To escape from the Earth's gravitational potential, acceleration of charged particles is needed. However, acceleration mechanism(s) remain unclear. In addition, the population of charged particles in the magnetosphere in the energy range of $\sim 1 - 100$\,keV has not been well studied observationally. Observing such charged particles is essential for understanding not only the mechanisms driving outflows but also the overall ecosystem of the Earth’s magnetosphere.

KSAT3-X is a CubeSat mission planned for launch in 2027, developed by Kanazawa University\footnote{\url{https://arc-sat.w3.kanazawa-u.ac.jp/en/research/research-01/}}, aimed to observe inflows and outflows of charged particles in the Earth's magnetosphere, particularly electrons and protons in the $10 - 40$\,keV energy range. The population of charged particles in this energy range has not yet been observed well. In fact, the reliable datasets of charged particles below 40\,keV are not available in the standard platform used to infer space radiation environments known as SPENVIS \cite{Heynderickx03, Kruglanski09, Heynderickx12}. The KSAT3-X mission is also expected to provide us with valuable information regarding the environments of charged particles that contribute to non-X-ray background events for future satellite missions, such as HiZ-GUNDAM \cite{yonetoku2025concept}.

We have developed radiation detectors based on diamonds as the mission instrument for KSAT3-X. KSAT3-X is a 3U-size CubeSat ($\sim10 \times 10 \times 30\,\mathrm{cm^3}$), of which $\sim 1$U ($\sim10 \times 10 \times 10\,\mathrm{cm^3}$) is allocated to the mission payload. KSAT3-X will be placed in a Sun-synchronous orbit at an altitude of $400-600\,\mathrm{km}$, and an expected mission lifetime is six months to one year. Because the volume available for the mission payload is limited, a compact instrument is required. Neither an active cooling system nor an onboard calibration source is preferred. A diamond radiation detector has the potential to meet these requirements for KSAT3-X. In addition, diamond is known to have high radiation tolerance \citep[e.g.,][]{Adam00}, meaning that its performance may remain stable in orbit. However, the radiation tolerance of diamond radiation detectors under the proton environment in orbit has not been well understood.

We have established an in-house production system of diamond used for a semiconductor at Kanazawa University \citep[e.g.,][]{Ichikawa24}. This system allows us to customize and improve not only the properties of diamonds but also the design of our diamond radiation detectors such as the electrode configuration. Furthermore, in-house production is important for the rapid and timely development of diamond radiation detectors. Therefore, it is necessary to compare the performance of our diamonds with that of commercially available ones.

In this paper, we evaluate the radiation tolerance of our diamond radiation detectors under conditions of our space applications by irradiating with 100\MeV\ protons at the Wakasa-wan Energy Research Center. We also measure the X-ray spectroscopic performance of our diamond radiation detectors to investigate their charge-collection performance. The spectroscopic performance is also used as an indicator of radiation tolerance. The paper is organized as follows. Section~\ref{sec:Adv_diamond} provides a brief overview of the advantages of diamonds for space use. Section~\ref{sec:lab} describes our diamond radiation detectors, the experimental setups, and their X-ray spectroscopic performance. Section~\ref{subsec:Radiation_tolerance_experiments} presents the radiation tolerance experiments. In Section~\ref{subsec:Discussion}, we discuss the obtained results and their implications. Finally, the summary and future prospects are summarized in Section~\ref{subsec:Summary_future}.

\section{Advantages of diamond for space use}
\label{sec:Adv_diamond}

Diamond has been widely using as a semiconductor owing to its remarkable physical properties. Single-crystal diamonds synthesized by chemical vapor deposition (CVD) method have a wider band gap and higher carrier mobility at room temperature than commonly used semiconductor materials (Table~\ref{tab:properties}).

\begin{table*}[ht]
  \caption{Key physical properties of diamond, silicon, germanium, and cadmium telluride used as semiconductors}
  \label{tab:properties}
  \centering
  \begin{tabular} {ll|c|c|c|c}
    \hline
    && Diamond & Si & Ge & CdTe\\
    \hline\hline
    Atomic number && 6 & 14 & 32 & 48, 52\\
    \hline
    Density ($\mathrm{g\, 
     cm^{-3}}$) && 3.52\,$^{\text{\cite{10.1063/1.1713740}}}$ & 2.33\,$^{\text{\cite{s90503491}}}$ & 5.33\,$^{\text{\cite{s90503491}}}$ & 6.20\,$^{\text{\cite{s90503491}}}$ \\
    \hline
    Band gap at 300\,K (eV) && 5.47\,$^{\text{\cite{Sze2006}}}$ & 1.12\,$^{\text{\cite{Sze2006}}}$ & 0.66\,$^{\text{\cite{Sze2006}}}$ & 1.56\,$^{\text{\cite{Sze2006}}}$ \\
    \hline
    \multirow{2}{*}{Mobility at 300\,K ($\mathrm{cm^2\,V^{-1}\, s^{-1}}$)} 
      & electron & 2200--4500\,$^{\text{\cite{doi:10.1126/science.1074374, Gabrysch11, Kholili24, Rahman26}}}$ & 1450\,$^{\text{\cite{Sze2006}}}$ & 3900\,$^{\text{\cite{Sze2006}}}$ & 1050\,$^{\text{\cite{Sze2006}}}$ \\
      & hole & 2300--3800\,$^{\text{\cite{doi:10.1126/science.1074374, Gabrysch11, Kholili24, Rahman26}}}$ & 500\,$^{\text{\cite{Sze2006}}}$ & 1900\,$^{\text{\cite{Sze2006}}}$ & 100\,$^{\text{\cite{Sze2006}}}$ \\
    \hline
    e-h pair creation energy (eV) && 13.1\,$^{\text{\cite{https://doi.org/10.1002/pssa.201600195}}}$\,-\,13.9\,$^{\text{\cite{KRAUS2021164947}}}$ & 3.62\,$^{\text{\cite{OWENS200418}}}$ & 2.96\,$^{\text{\cite{OWENS200418}}}$ & 4.43\,$^{\text{\cite{OWENS200418}}}$ \\
    \hline
    Dielectric constant && 5.7\,$^{\text{\cite{Sze2006}}}$ & 11.9\,$^{\text{\cite{Sze2006}}}$ & 16.0\,$^{\text{\cite{Sze2006}}}$ & 10.2\,$^{\text{\cite{Sze2006}}}$ \\
    \hline
  \end{tabular}
\end{table*}

The characteristics of CVD diamonds exhibit their great potential for space applications \cite{Pace03}. For instance, CVD diamond is insensitive to visible light because of its band gap of 5.47\,eV (i.e., solar blindness) \cite{Sze2006}. In the early stages, diamond detectors were used as ultraviolet detectors for solar physics \citep[e.g.,][]{Hochedez00, Hochedez01, Benmoussa03, DeSio03, Benmoussa06, Hochedez06, Mendoza15}. In addition to such application, CVD diamond is considered as a promising mission instrument onboard a CubeSat. This is because (1) the leakage current (also known as the dark current) of a CVD diamond is extremely low at room temperature even under high bias voltages, (2) no optical blocking filter is needed because of the solar-blindness, and (3) a CVD diamond can operate normally not only at room temperature but also under both higher and lower temperature conditions. Therefore, a diamond radiation detector is a strong candidate for the mission instrument onboard KSAT3-X.

\section{Laboratory experiments}
\label{sec:lab}

\subsection{Our diamond radiation detectors}
\label{sec:our_diamonds}

\begin{figure*}[ht]
	\centering
	\includegraphics[width=140mm]{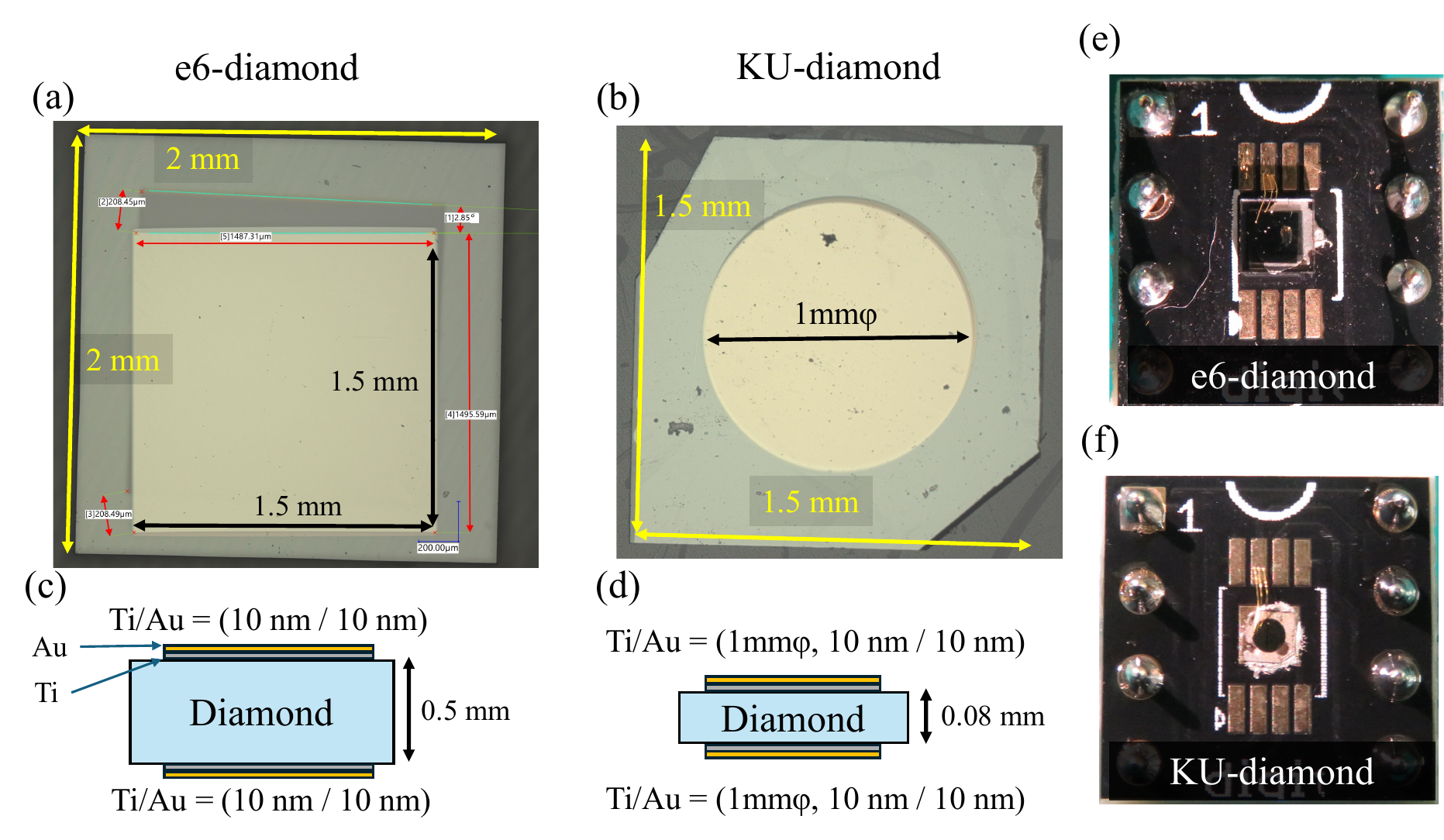}
	\caption{(a) Micrograph of \ESix\ with the metal electrodes. (b) Same as (a), but for \KU. (c) Schematic cross-sectional view of \ESix. (d) Same as (c), but for \KU. (e) Photograph of \ESix\ mounted on a circuit board. (f) Same as (e), but for \KU.}
	\label{fig:Two_diamonds}
\end{figure*}

We have developed CVD diamond-based radiation detectors as mission instruments for KSAT3-X, aimed at observing low-energy charged particles in the Earth's magnetosphere. For this purpose, we investigate the performance of two types of CVD diamond. The first is a commercial, high-quality microwave plasma CVD (MPCVD) single-crystal diamond manufactured by Element Six (hereafter \ESix), and the second is an MPCVD single-crystal diamond fabricated in-house at Kanazawa University (hereafter \KU). For both diamonds, metal electrodes are deposited to readout signal charges. The specifications of the two diamond radiation detectors are summarized in Table~\ref{tab:Two_diamonds}. Fig.~\ref{fig:Two_diamonds} shows the pictures of the two diamond radiation detectors.

As mentioned above, we have established an in-house production system for developing CVD diamonds at Kanazawa University \citep[e.g.,][]{Ichikawa24}, enabling us to customize the design of the diamond radiation detectors. To fabricate \KU, we conduct growth experiments using an MPCVD system equipped with a load-lock sample-exchange chamber\footnote{Arios Inc., Japan}. We grow a 90\,$\mu$m diamond layer on a high-pressure, high-temperature (HPHT) type-Ib diamond (100) substrate ($2 \times 2 \times 0.3$\,mm). The growth conditions are as follows: (1) a power input of 1500\,W, (2) a pressure of 45\,kPa, (3) a methane concentration (CH$_{4}$/H$_{2}$) of 5\,\%, and (4) a substrate temperature of 1200\,$^{\circ}$C. After the growth, we obtain an 80\,$\mu$m-thick, free-standing CVD diamond plate through laser cutting and polishing. Although batch-to-batch variations may lead to slightly higher contamination levels in the sample used for the measurements, secondary ion mass spectrometry (SIMS) measurements on a sample grown under the same conditions show that boron and nitrogen contaminations are below the detection limits, i.e., B $<$ 0.006\,ppm and N $<$ 0.1\,ppm, respectively. 

We deposit a metal electrode with a thickness of 20\,nm (Ti 10\,nm \& Au 10\,nm, see Fig.~\ref{fig:Two_diamonds}) on the diamond, which is thinner than those used in previous studies \citep[typically $150 - 200$\,nm, e.g.,][]{Pomorski06, Shimaoka16, Shimaoka21}. Such a thin metal electrode is essential to improve the detectability of low energy charged particles. To achieve this configuration, after the growth of the diamond, we perform oxygen termination using a mixed acid solution of HNO$_{3}$ and H$_{2}$SO$_{4}$ ($1:3$). Then, the diamond is heated to 200\,$^{\circ}$C. Finally, we deposit two metal electrodes of Ti 10\,nm and Au 10\,nm on each surface of the diamond by vacuum evaporation.

\begin{table*}[htpb]
  \caption{Specifications of the two diamond radiation detectors used in this paper.} 
  \label{tab:Two_diamonds}
  \centering
  \begin{tabular} {l|c|c}
    \hline
    Name & \ESix & \KU \\
    \hline\hline
    Manufacturer & Element Six & Kanazawa University \\
    \hline
    Manufacturing method& Microwave plasma CVD & Microwave plasma CVD \\
    \hline
    Surface area ($\mathrm{mm^2}$) & $2.0 \times \, 2.0 $ & $1.5 \times \, 1.5$ \\
    \hline
    Thickness ($\mathrm{mm}$) & 0.5 & 0.08\\
    \hline
    Electrode & Ti (10\,nm) / Au (10\,nm) & Ti (10\,nm) / Au (10\,nm)\\
    \hline
    Electrode area & $1.5 \times \, 1.5 \, \mathrm{mm^2}$ & $1 \, \mathrm{mm} \phi$\\
    \hline
  \end{tabular}
\end{table*}

\subsection{Experimental Setup}
\label{subsec:Ex_setup}

\begin{figure*}[h]
	\centering
	\includegraphics[width= 150mm]{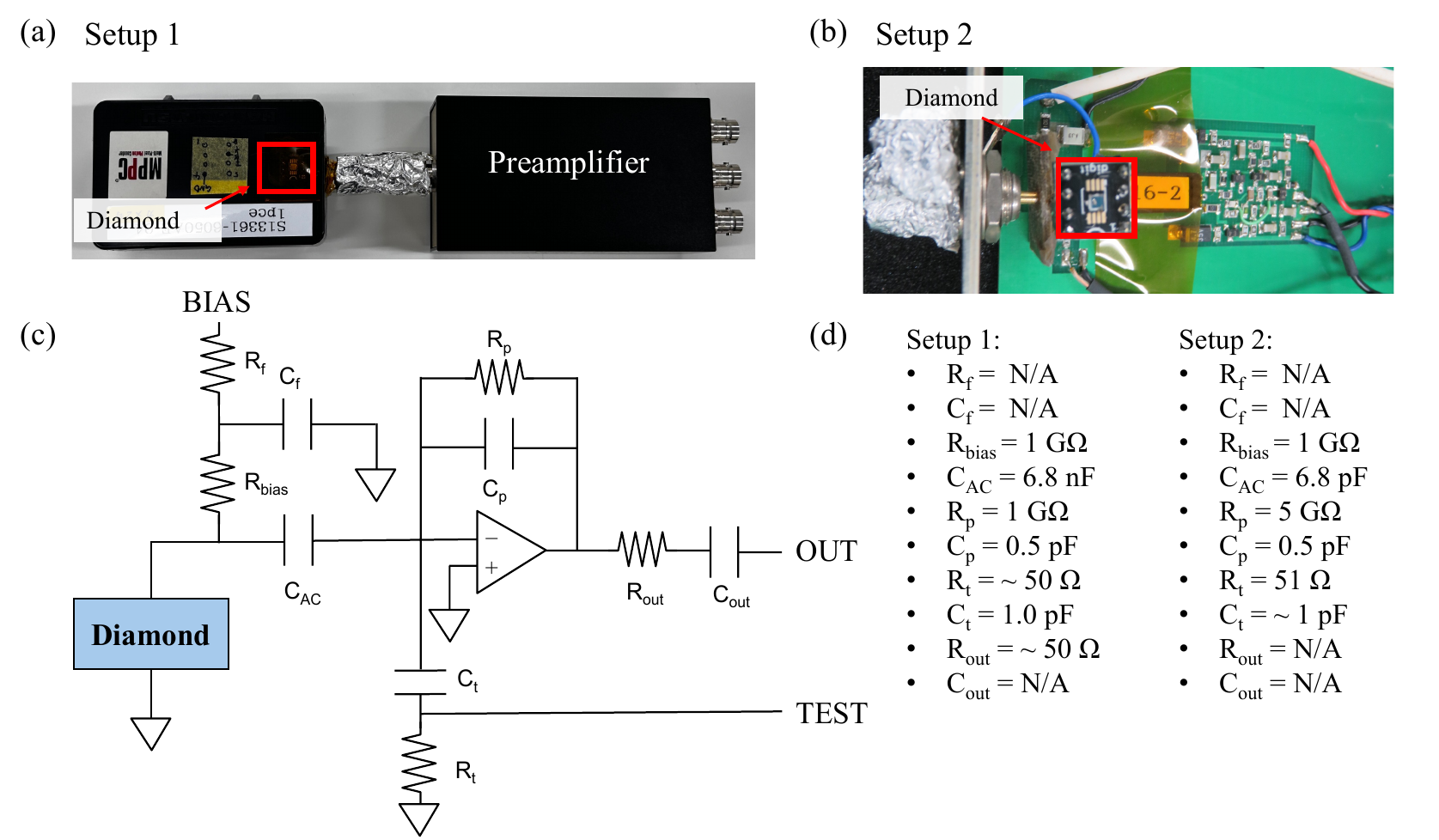}
	\caption{(a) Photograph of Setup\,1. (b) Same as (a), but for Setup\,2. (c) Circuit diagrams around the preamplifier. (d) Specification sheet of the circuit diagrams for Setup\,1 and Setup\,2. Unavailable values are indicated as NA.}
	\label{fig:setup}
\end{figure*}

We, here, employ two types of readout systems: Setup\,1 and Setup\,2 (see Fig.\,\ref{fig:setup}). Setup\,1 includes a preamplifier of CLEAR-PULSE CO., LTD. (CLEAR-PULSE 595L\footnote{\url{https://clearpulse.co.jp/en/ProductInfo_item_detail?index=32}}). The diamond is installed in a box connected to the preamplifier box. Setup\,2 is designed to directly mount the diamond onto electronic circuits closely connected to a custom preamplifier (equivalent to CLEAR-PULSE 580K\footnote{\url{https://clearpulse.co.jp/en/ProductInfo_item_detail?index=28}}). The reason of why we have the two systems is that Setup\,1 was initially used as the readout system; however, we found that its noise level was relatively high, which prevented measurements of X-ray spectra in the lower energy range below 20 keV. To address this issue, we developed Setup\,2 to reduce the noise level. In this paper, the performance of \ESix\ is evaluated using both Setup\,1 and Setup\,2, whereas that of \KU\ is investigated using Setup\,2 only.

We measure the spectroscopic performance of the two diamond radiation detectors at room temperature (typically $\sim 20$\,$^{\circ}$C) with a bias voltage of $+200$\,V. In the measurements, the detectors are operated in photon-counting mode. For Setup\,2, the shaping time constant of the shaper (see Fig.~\ref{fig:setup}) is set to 1\,$\mu$s and 3\,$\mu$s for the pre- and post examinations of the proton irradiation experiments, respectively, in order to optimize our experimental setups. DC regulated power supplies are used for the preamplifiers.

\subsection{X-ray spectroscopic performance} \label{subsec:Spectroscopic_performance}

We investigate the X-ray spectroscopic performance of the diamond radiation detectors using characteristic X-rays from three radioisotope sources: \Am\ ($\gamma$ ray: 59.5\keV), \Ba\ (Cs-K$\alpha$: 31.0\keV, Cs-K$\beta$: 35.0\keV), and \Cd\ (Ag-K$\alpha$: 22.2\keV, Ag-K$\beta$: 24.9\keV). We count a number of electrons as the signal charge produced through photoelectric absorption and the Compton scattering in the diamond. Finally, we acquire energy spectra with Amptek MultiChannel Analyzer (MCA).

\begin{figure*}[ht]
	\centering
	\includegraphics[width=0.9\linewidth]{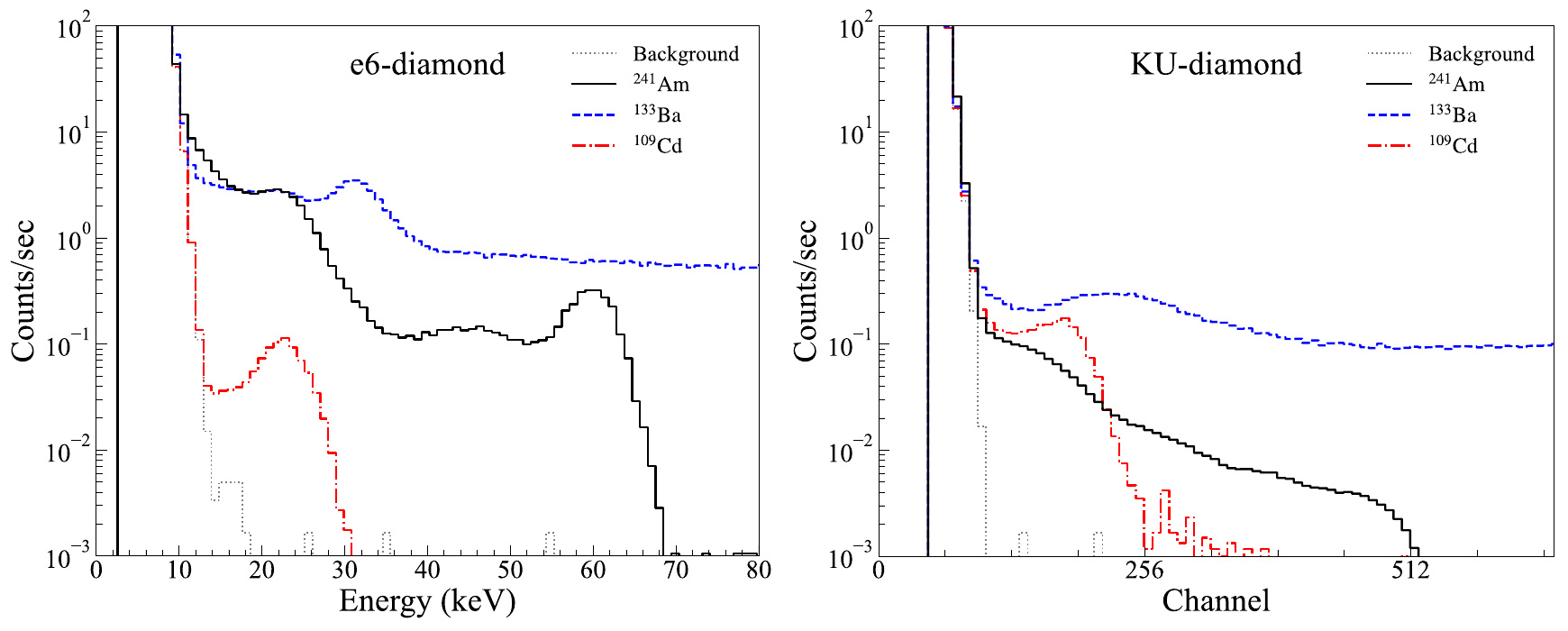}
	\caption{
	Observed energy spectra of \Am, \Ba, \Cd, and the background, respectively, for \ESix\ (left) and \KU\ (right). The black solid, blue dashed, red dash-dotted, and gray dotted lines represent the obtained spectra of  \Am, \Ba, \Cd, and the background, respectively, obtained with Setup\,2. The spectra are binned with 8\,channels for display purpose.
	}
	\label{fig:e6-c-100-27_and_s-1b1101-1_0yr}
\end{figure*}

The left panel of Fig.~\ref{fig:e6-c-100-27_and_s-1b1101-1_0yr} shows the X-ray spectra of the three radioisotope sources obtained with \ESix, respectively. The photoelectric peak of each source is clearly detected, indicating that the signal charge can be successfully acquired with the system. Then, we examine the linearity of \ESix. The linearity is defined as $y = ax + b$, where $x$ and $y$ denote the incident X-ray energy in units of keV and the observed pulse height amplitude in units of channel, respectively, and the best-fit values of $a$ and $b$ are measured at $a = 11.76 \pm 0.05$\,ch\,keV$^{-1}$ and $b = -\,40.72 \pm 2.70$\,ch, respectively. In addition, we measure the energy resolution (FWHM) of \ESix\ at $3.39 \pm 0.16$\,keV for \Am, $2.56^{+0.36}_{-0.31}$\,keV for \Ba, and $3.29^{+0.41}_{-0.82}$\,keV for \Cd. The contribution from the noise of the system is subtracted when measuring the energy resolution. For the \Ba\ spectra, an additional power-law component is included to describe the continuum component. The current energy resolution of \ESix\ is not able to resolve the characteristic X-rays of \Ba\ and \Cd\ into the K$\alpha$ and K$\beta$ lines. The background spectrum of the system indicates that X-rays above $\sim 12$\,keV are detectable with \ESix. In fact, although it is difficult to determine the peak position of the 13.9\,keV line from \Am, a spectral feature associated with the line can be identified on its higher-energy side.

We find that the spectroscopic performance of \KU\ is different from that of \ESix. As shown in the right panel of Fig.~\ref{fig:e6-c-100-27_and_s-1b1101-1_0yr}, the photoelectric peak at 59.5\keV\ is not detected in the observed spectrum. However, the photoelectric peaks corresponding to \Cd\ and \Ba, respectively, are detected. The reason of why \KU\ appears to exhibit a different spectroscopic performance will be discussed in Section~\ref{sec:diff}. Although it is hard to examine the linearity of \KU, the obtained background spectrum may indicate that X-rays above 100\,channels corresponding to $\sim 12$\,keV are detectable. Since the thickness of the diamond for \ESix\ is a factor of $\sim 6$ larger than that of \KU, the observed count rate is different between \ESix\ and \KU.

\section{Radiation tolerance experiments} \label{subsec:Radiation_tolerance_experiments}

We investigate the radiation tolerance of our diamonds, which is one of the important parameters for space applications. To this end, we conduct proton irradiation experiments on \ESix\ and \KU\ using 100\,MeV protons at the Wakasa-wan Energy Research Center.

Based on the configuration of the KSAT3-X mission, we infer the proton environment in its orbit. We first assume that the diamond is shielded by the detector housing and satellite body made of aluminium with a cumulative thickness of 11.1\,mm. Next, an altitude of 600\,km is assumed for KSAT3-X in a Sun-synchronous orbit. Then, we calculate the expected proton environment using the SHIELDOSE-2 application implemented in SPENVIS \cite{Heynderickx03, Kruglanski09, Heynderickx12}. For the dose estimate, contributions from trapped protons, electrons, and bremsstrahlung are taken into account. In the calculation, due to the limitations of SPENVIS, we assume that the diamond is exposed to protons in the energy range from 40\,keV to 1000\,MeV. Thus, we find that 100\,MeV protons are dominant in the assumed orbital environments, indicating that the diamond is expected to be mainly exposed to 100\,MeV protons. To estimate the total dose rate deposited in the diamond by 100\,MeV protons, we use PSTAR \cite{Berger92}. Since PSTAR does not include the physical parameters of diamond, we instead adopt amorphous carbon ($\rho = 2.0$\,$\mathrm{g\,cm}^{-3}$), which has a density comparable to that of diamond ($\rho = 3.52$\,$\mathrm{g\,cm}^{-3}$). Note that, for materials composed of the same element, the mass stopping power in units of MeV\,cm$^{2}$\,g$^{-1}$ depends only weakly on the material density and bonding structure. Therefore, the difference between amorphous carbon and diamond is expected to be small. Although we calculate the total dose rate provided by protons over the full energy range from 40\,keV to 1000\,MeV, we treat 100\,MeV protons as representative of the proton environment. Specifically, the total number of protons to be irradiated to the diamond is estimated based on the ratio of the total dose rate from all protons to that deposited by a single 100\,MeV proton. This assumption is justified because of the fact that 100\,MeV protons are the most dominant component in the environments considered. Although 100\,MeV protons are adopted as a representative of the proton environment, contributions from protons with energies below 100\,MeV are taken into account in the dose calculation. Finally, we obtain the fluence of 100\,MeV protons in the orbit based on the ionizing energy loss (IEL) as $8.89 \times 10^{8}\,\mathrm{protons\,cm^{-2}\,yr^{-1}}$. We also estimate an equivalent fluence of 100\,MeV protons in the orbit based on the non-ionizing energy loss (NIEL) at $7.79 \times 10^{8}\,\mathrm{protons\,cm^{-2}\,yr^{-1}}$, which is comparable to that estimated for the IEL. Here, we adopt a more conservative estimate for the required fluence of 100\,MeV protons.

We irradiate the diamonds with cumulative fluences of 100\,MeV protons equivalent to 3\,months, 1\,year, 3\,years, and 10\,years in the orbit, respectively. We additionally irradiate \ESix\ with cumulative fluences of 100\,MeV protons equivalent to 100\,years because, as discussed later, no degradation of the spectroscopic performance is observed for \ESix\ after irradiation equivalent to 10\,years. The motivation for extending the proton irradiation beyond the expected mission lifetime of KSAT3-X is to determine the onset of performance degradation in diamond detectors. Such information is important for understanding the detector physics of diamonds and long-term reliability of diamond detectors.

\subsection{Radiation tolerance of \ESix}

We evaluate the radiation tolerance of \ESix\ based on its spectroscopic performance, i.e., its linearity and energy resolution. Since only Setup\,1 enables us to perform consistent measurement and comparison for the linearity, we measure the linearity using Setup\,1. This is because the configuration of Setup\,2 was optimized at a later stage after the 10-year equivalent irradiation, it is hard to fairly compare the linearity across different epochs. For measurements of the energy resolution, Setup\,2 is used to maximize the performance. The contribution from the noise of the system is subtracted, allowing us to measure the intrinsic energy resolution of the diamond.

\begin{figure*}[ht]
	\centering
	\includegraphics[width=0.9\linewidth]{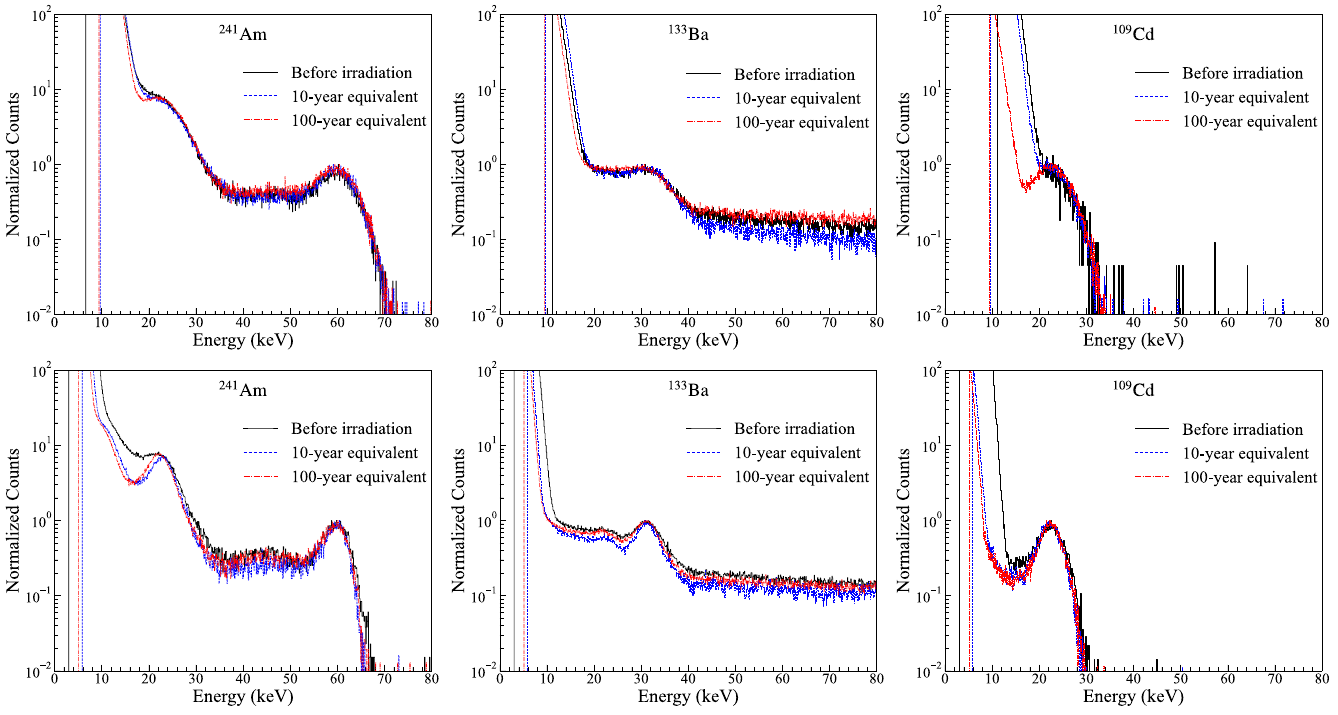}
	\caption{Top: energy spectra of \Am\ (left), \Ba\ (middle), and \Cd\ (right), respectively, obtained with Setup\,1 for \ESix. The black solid, blue dashed, and red dash-dotted lines represent the energy spectra obtained before irradiation, after the 10-year equivalent irradiation, and the 100-year equivalent irradiation, respectively, for each radioisotope source. The photon count of the photoelectric peak of each source is normalized to unity for a comparison purpose. Bottom: same as the top panels, but for those obtained with Setup\,2. 
	}
	\label{fig:e6-c-100-27_0-100yr}
\end{figure*}

We find no degradation in the spectroscopic performance of \ESix\ before and after the proton irradiation experiments, even for irradiation equivalent to 100\,years. Fig.~\ref{fig:e6-c-100-27_0-100yr} shows the energy spectra of the three radioisotope sources obtained with \ESix\ using Setup\,1 and Setup\,2 before and after the proton irradiation experiments, respectively. Even though the background level of the system varies among measurements, the overall spectral shapes including the photoelectric peaks appear consistent with each other, indicating that no apparent radiation-induced damage is observed. In fact, the leakage current remains unchanged after the proton irradiation experiments. Since the charge collection efficiency (or charge collection distance) is also used as an indicator to investigate the spectroscopic performance, we will evaluate these parameters in a future study.

\begin{figure*}[ht]
	\centering
	\includegraphics[width=0.9\linewidth]{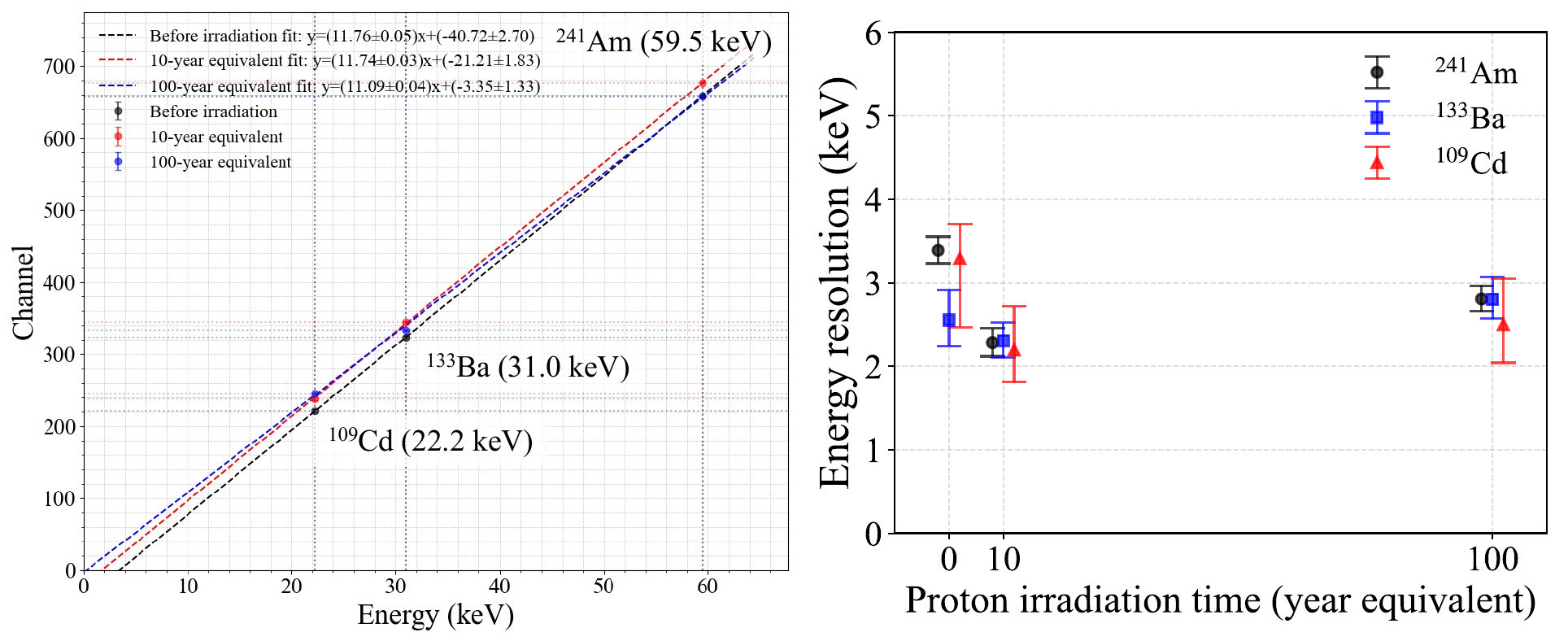}
	\caption{Left: linearity of \ESix\ measured before and after the proton irradiation experiments. The black, red, and blue data points represent the line centroids of each source obtained before irradiation, the 10-year, and the 100-year equivalent irradiation, respectively. Right: same as the left panel, but for the energy resolution. The black circle, blue box, and red triangle data points show the trends of the observed energy resolution for \Am, \Ba, and \Cd, respectively. 
	}
	\label{fig:FWHM_linearity}
\end{figure*}

The left panel of Fig.~\ref{fig:FWHM_linearity} shows the linearity before and after the proton irradiation experiments. Since the measurements were performed with Setup\,1, the two lines from \Ba\ and \Cd\ are not resolved and can be modeled with a single Gaussian component. Therefore, only three data points are used to evaluate the linearity. The linearity obtained after irradiation equivalent to 10\,years is consistent with that derived before irradiation, namely, the slope $a$ is measured at $a = 11.76 \pm 0.05$ for the initial \ESix\ and $a = 11.74 \pm 0.03$ after the 10-year equivalent irradiation. After the 100-year equivalent irradiation, the slope shows a slight decrease of $\sim 6$\,\%, i.e., $a = 11.09 \pm 0.04$. Nevertheless, \ESix\ exhibits almost the same linearity before and after the proton irradiation experiments.

The energy resolution of \ESix\ remains unchanged, as shown in the right panel of Fig.~\ref{fig:FWHM_linearity}. As mentioned above, the contribution from the noise of the system has been subtracted in the analysis. In addition, we apply a double Gaussian model to the observed energy spectra of \Ba\ and \Cd\ obtained with Setup\,2. Although the K$\alpha$ and K$\beta$ lines are not fully resolved, the photoelectric peaks exhibit an asymmetric shape toward higher energies, indicating that the contribution from the K$\beta$ line may be measurable. Therefore, in the fitting, we set that (1) the width of the K$\beta$ line is tied to that of the K$\alpha$ line, (2) the ratio of the total photon counts between the K$\alpha$ and K$\beta$ components is fixed at that of their emission probabilities, and (3) the line centroid for the K$\beta$ line is constrained to be higher than that of the K$\alpha$ line. The resulting energy resolution of \ESix\ as a function of irradiation epoch is shown in the right panel of Fig.~\ref{fig:FWHM_linearity}.

\subsection{Radiation tolerance of \KU}

\begin{figure*}[ht]
	\centering
	\includegraphics[width=0.99\linewidth]{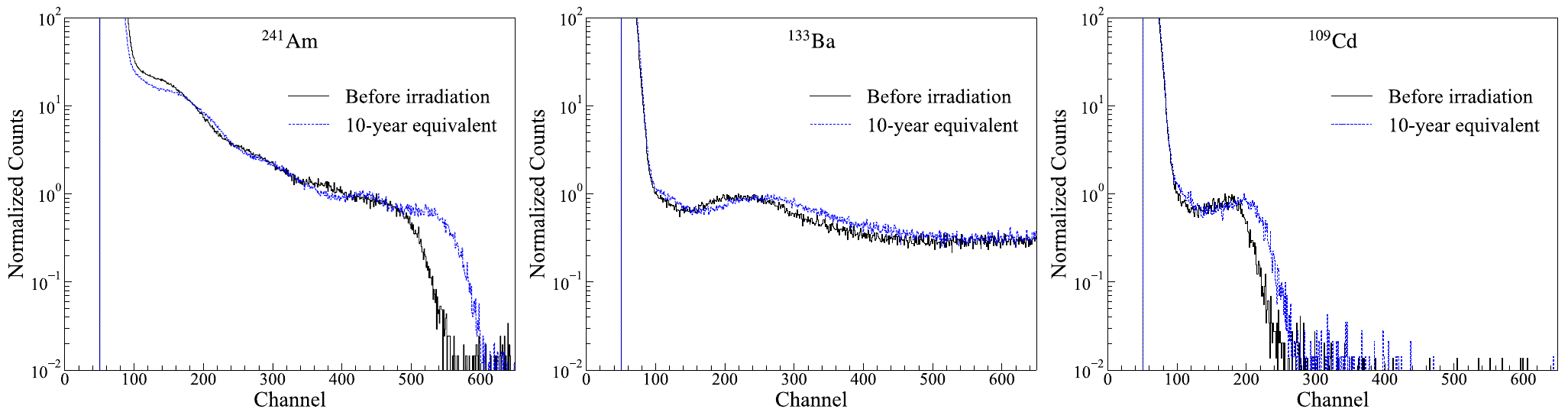}
	\caption{Same as Fig.~\ref{fig:e6-c-100-27_0-100yr}, but for \KU. The energy spectra are obtained with Setup\,2. The black solid and blue dotted lines describe the energy spectra obtained before and after the proton irradiation experiments. 
	}
	\label{fig:s-1b1101-1_0-10yr}
\end{figure*}

We, here, examine the radiation tolerance of \KU. As mentioned in Section~\ref{subsec:Spectroscopic_performance}, no photoelectric peak of \Am\ has been observed before irradiation. Even after the 10-year equivalent irradiation, no photoelectric peak of \Am\ is detected in the obtained energy spectrum (see the left panel of Fig.~\ref{fig:s-1b1101-1_0-10yr}). However, for \Ba\ and \Cd, we observe the photoelectric peaks even after the 10-year equivalent irradiation (see the middle and right panels of Fig.~\ref{fig:s-1b1101-1_0-10yr}). These results indicate no significant degradation in the spectroscopic performance of \KU.

\section{Discussion}
\label{subsec:Discussion}

We have investigated the spectroscopic performance of \ESix\ and \KU\ before and after the proton irradiation experiments. Based on these results, we discuss the radiation tolerance of the diamonds for space applications. In addition, we discuss possible physical origins of the observed difference in spectroscopic performance between \ESix\ and \KU.

\subsection{Radiation tolerance of the diamonds for space use}
\label{sec:tolerance}

\ESix\ exhibits no degradation in the spectroscopic performance before and after the proton irradiation experiments up to the 100-year equivalent, indicating that \ESix\ has high radiation tolerance under the assumed environmental conditions. Although the linearity of \ESix\ decreases by $5.7_{-0.5}^{+0.5}\,\%$ after the 100-year equivalent irradiation, the performance of \ESix\ can remain unchanged over the expected mission lifetime of KSTA3-X. In fact, the energy resolution of \ESix\ is consistent across all epochs (see the right panel of Fig.~\ref{fig:FWHM_linearity}). Since the diamond shielded by aluminium with a cumulative thickness of 11\,mm is feasible for a CubeSat, we conclude that a diamond radiation detector based on \ESix\ is suitable for space use. No degradation in the spectroscopic performance indicates that no onboard calibration source is required. Because of the radiation tolerance, compactness, solar-blindness, and working temperature, the \ESix-based diamond radiation detector allows us to use it as the mission instrument onboard KSAT3-X.

In addition to \ESix, \KU\ is suitable for use as the mission instrument onboard KSAT3-X. Even though the photoelectric peak of \Am\ is not observed in the energy spectrum, those of \Ba\ and \Cd\ are detected (see Fig.~\ref{fig:s-1b1101-1_0-10yr}), suggesting that \KU\ is capable of measuring signal charges at least up to 35\,keV. This level of spectroscopic performance meets the requirements for the KSAT3-X mission. In addition, no apparent degradation in the spectroscopic performance of \KU\ has been found after proton irradiation. In terms of radiation tolerance, \KU\ has a similar performance to \ESix\ at least up to the 10-year equivalent irradiation. Therefore, \KU\ is also a candidate for the mission instrument onboard KSAT3-X.

\subsection{Difference in spectroscopic performance between \ESix\ and \KU}
\label{sec:diff}

We, here, discuss possible physical origins of the observed difference in spectroscopic performance between \ESix\ and \KU. As mentioned in Section~\ref{subsec:Spectroscopic_performance}, \ESix\ has a higher energy resolution than \KU.

One possible explanation for the observed difference is that \KU\ has a factor of $\sim 6$ smaller thickness of diamond than \ESix, leading to (1) higher capacitance of \KU, causing higher noise and/or (2) smaller hit efficiency of high-energy photons due to the low mass absorption coefficient of diamond at high energy. The surface area of \KU\ is comparable to that of \ESix\ (see Table~\ref{tab:Two_diamonds}), and the electrode areas of the two detectors are also similar. The thickness is the only significant difference in the geometric parameters of the diamond. Thus, the ratio of thickness to surface area may play an important role in determining spectroscopic performance. This hypothesis can be tested using a thicker KU-type diamond as well as a thinner E6-type diamond.

Another possible explanation is that the spectral response of \ESix\ has a smaller low-energy tail component than \KU. As analog to silicon-based detectors such as charge-coupled devices (CCDs), e.g., \cite{Janesick01, Ueda13, Inoue16}, it is possible to consider that a low-energy tail component is generated due to partial absorption of signal charges inside the diamond. In fact, such a component is observed in both diamonds. For instance, the energy spectrum of \Am\ obtained with \ESix\ shows a low-energy tail component below 50\,keV (see the bottom left panel of Fig.~\ref{fig:e6-c-100-27_0-100yr}). Therefore, the absence of a detectable photoelectric peak of \Am\ with \KU\ may be attributed to a larger degree of partial charge absorption within the diamond.

As a possible origin of partial charge absorption, defects in the surface layer of the diamond may partially trap signal charges. Polishing is needed to conduct chemical surface treatments and may cause damage extending to depths of several tens of micrometers. Such damage may contribute weakly to \ESix\ because of its thickness of $\sim 500$\,$\mu$m, whereas it may be significant for \KU\ (its thickness of $\sim 80$\,$\mu$m). The impact of surface-layer conditions on charge trapping can be investigated using a thicker KU-type diamond as well as a thinner E6-type diamond. Although the effect of charge traps in the diamond bulk may be reduced by pumping the crystal with photon or electron irradiation prior to the measurements, we did not apply this technique in this study. Further investigation is needed to quantify the impact of this effect on the detector performance. Note that the typical exposure time for the \KU\ measurements exceeds 100\,min.

Another possibility is that bulk defect/trap densities and/or contact interface states contribute to partial charge collection. Although the nitrogen contamination levels in both diamonds are comparable to or below the SIMS detection limit (N $< 0.1$\,ppm), nitrogen-related centers and other deep levels may primarily act as trapping and recombination sites, thereby enhancing the low-energy tail component. In addition, a rectifying (Schottky) behavior can be expected at the interface between a Ti-electrode and diamond for high-resistivity diamond. Then, variations in surface termination and interface-state density may influence the effective barrier and near-contact charge collection. These possibilities should be quantified by dedicated electrical and transient measurements. Furthermore, these possibilities can be tested by comparing electron- and hole-collection configurations and/or using a highly conductive p-type (boron-doped) contact layer. In conclusion, the spectral response of diamond radiation detectors, particularly in the energy range of $1 - 100$\,keV, has not yet been well investigated. Therefore, further studies are required to arrive at a firm conclusion regarding the hypothesis and these possibilities.

\section{Summary and future prospects}\label{subsec:Summary_future}

In this paper, we have first investigated the X-ray spectroscopic performance of the two diamond radiation detectors, \ESix\ and \KU, using three radioisotope sources (\Am, \Ba, and \Cd) to evaluate their charge-collection performance. Then, we have conducted proton irradiation experiments with 100\,MeV protons at the Wakasa-wan Energy Research Center to examine the radiation tolerance of the diamonds. The goal of this study was to evaluate whether or not these diamonds can be used as the mission instrument onboard KSAT3-X. The main conclusions of this paper are summarized as follows:

\begin{enumerate}

\item Both \ESix\ and \KU\ successfully detect the characteristic X-rays from \Ba\ and \Cd, respectively, whereas only \ESix\ resolves the 59.5\,keV $\gamma$ ray from \Am.

\item The radiation tolerance of both diamonds is sufficiently high up to at least the 10-year equivalent irradiation under the assumed orbital conditions, where the diamond is shielded by aluminium with a cumulative thickness of 11.1\,mm. We additionally irradiated \ESix\ with 100\,MeV protons up to the 100-year equivalent. As a result, no significant degradation in the spectroscopic performance is observed. 

\item Both \ESix\ and \KU\ can be used as the mission instrument onboard KSAT3-X. Even though \ESix\ has a higher energy resolution than \KU, the current spectroscopic performance of \KU\ meets the requirements of the KSAT3-X mission.

\end{enumerate}

We still have room for improvement in developing the diamond radiation detectors. One of our plans is to investigate the spectral response of the diamonds in the energy range of $1 - 100$\,keV. This is because we need to understand the physics of diamond radiation detectors in detail. There is a lack of information to interpret the observed energy spectra of the three radioisotope sources thus far. For instance, the level of nitrogen contamination is an important factor. In addition, the fabrication of boron-doped diamonds is crucial for revealing the physics of diamond radiation detectors. Another important step is to irradiate the diamond with charged particles. Since the KSAT3-X mission is aimed to observe charged particles in the Earth's magnetosphere, we need to investigate the capability of measuring charged particles directly. Therefore, the in-house production for fabricating CVD diamonds at Kanazawa University is vital for customizing and optimizing the performance of diamond radiation detectors. We will be able to develop improved diamond radiation detectors based on a wide range of ideas and hypotheses.

\section*{Acknowledgments}
We are grateful to the anonymous referees for helpful suggestions and comments.
We acknowledge support by Program for Forming Japan's Peak Research Universities (J-PEAKS) Grant Number JPJS00420230006. 
This work was supported by JSPS KAKENHI grant numbers JP25K23398 and 	JP26K00741(S.U.).

\bibliographystyle{unsrt}

\bibliography{cas-refs.bib}

@article{Yau1997,
  title = {Sources of ion outflow in the high latitude ionosphere},
  author = {Yau, A. W. and Andre, M.},
  journal = {Space Science Reviews},
  volume = {80},
  number = {1},
  pages = {1--25},
  year = {1997},
  doi = {10.1023/A:1004947203046},
  url = {https://doi.org/10.1023/A:1004947203046}
}

@article{yonetoku2025concept,
  title={Concept of high-z gamma-ray bursts unraveling the dark ages and extreme space-time mission—HiZ-GUNDAM},
  author={Yonetoku, Daisuke and Doi, Akihiro and Mihara, Tatehiro and Matsuhara, Hideo and Sakamoto, Takanori and Tsumura, Kohji and Ioka, Kunihito and Arimoto, Makoto and Ando, Yoshiyuki and Enoto, Teruaki and others},
  journal={Journal of Astronomical Telescopes, Instruments, and Systems},
  volume={11},
  number={4},
  pages={044002--044002},
  year={2025},
  publisher={Society of Photo-Optical Instrumentation Engineers}
}

@book{Sze2006,
  title     = {Physics of Semiconductor Devices},
  author    = {Sze, S. M. and Ng, Kwok K.},
  edition   = {3rd},
  year      = {2006},
  publisher = {Wiley-Interscience},
  address   = {Hoboken, N.J.}
}

@article{
doi:10.1126/science.1074374,
author = {Jan Isberg  and Johan Hammersberg  and Erik Johansson  and Tobias Wikstrom  and Daniel J. Twitchen  and Andrew J. Whitehead  and Steven E. Coe  and Geoffrey A. Scarsbrook },
title = {High Carrier Mobility in Single-Crystal Plasma-Deposited Diamond},
journal = {Science},
volume = {297},
number = {5587},
pages = {1670-1672},
year = {2002},
doi = {10.1126/science.1074374},
URL = {https://www.science.org/doi/abs/10.1126/science.1074374},
eprint = {https://www.science.org/doi/pdf/10.1126/science.1074374}}

@article{KRAUS2021164947,
title = {Charge carrier properties of single-crystal CVD diamond up to 473 K},
journal = {Nuclear Instruments and Methods in Physics Research Section A: Accelerators, Spectrometers, Detectors and Associated Equipment},
volume = {989},
pages = {164947},
year = {2021},
issn = {0168-9002},
doi = {https://doi.org/10.1016/j.nima.2020.164947},
url = {https://www.sciencedirect.com/science/article/pii/S0168900220313449},
author = {Benjamin Kraus and Patrick Steinegger and Nikolay V. Aksenov and Rugard Dressler and Robert Eichler and Erich Griesmayer and Dominik Herrmann and Andreas Turler and Christina Weiss}
}

@article{https://doi.org/10.1002/pssa.201600195,
author = {Shimaoka, Takehiro and Kaneko, Junichi H. and Sato, Yuki and Tsubota, Masakatsu and Shimmyo, Hiroaki and Chayahara, Akiyoshi and Watanabe, Hideyuki and Umezawa, Hitoshi and Mokuno, Yoshiaki},
title = {Fano factor evaluation of diamond detectors for alpha particles},
journal = {physica status solidi (a)},
volume = {213},
number = {10},
pages = {2629-2633},
doi = {https://doi.org/10.1002/pssa.201600195},
url = {https://onlinelibrary.wiley.com/doi/abs/10.1002/pssa.201600195},
eprint = {https://onlinelibrary.wiley.com/doi/pdf/10.1002/pssa.201600195},
year = {2016}
}

@article{OWENS200418,
title = {Compound semiconductor radiation detectors},
journal = {Nuclear Instruments and Methods in Physics Research Section A: Accelerators, Spectrometers, Detectors and Associated Equipment},
volume = {531},
number = {1},
pages = {18-37},
year = {2004},
note = {Proceedings of the 5th International Workshop on Radiation Imaging Detectors},
issn = {0168-9002},
doi = {https://doi.org/10.1016/j.nima.2004.05.071},
url = {https://www.sciencedirect.com/science/article/pii/S0168900204010575},
author = {Alan Owens and A. Peacock}
}

@article{10.1063/1.1713740,
    author = {Mykolajewycz, R. and Kalnajs, J. and Smakula, A.},
    title = {High‐Precision Density Determination of Natural Diamonds},
    journal = {Journal of Applied Physics},
    volume = {35},
    number = {6},
    pages = {1773-1778},
    year = {1964},
    month = {06},
    issn = {0021-8979},
    doi = {10.1063/1.1713740},
    url = {https://doi.org/10.1063/1.1713740},
    eprint = {https://pubs.aip.org/aip/jap/article-pdf/35/6/1773/18332135/1773_1_online.pdf},
}

@Article{s90503491,
AUTHOR = {Del Sordo, Stefano and Abbene, Leonardo and Caroli, Ezio and Mancini, Anna Maria and Zappettini, Andrea and Ubertini, Pietro},
TITLE = {Progress in the Development of CdTe and CdZnTe Semiconductor Radiation Detectors for Astrophysical and Medical Applications},
JOURNAL = {Sensors},
VOLUME = {9},
YEAR = {2009},
NUMBER = {5},
PAGES = {3491--3526},
URL = {https://www.mdpi.com/1424-8220/9/5/3491},
PubMedID = {22412323},
ISSN = {1424-8220},
DOI = {10.3390/s90503491}
}

@Article{Shelley72,
  author   = {{Shelley}, E.~G. and {Johnson}, R.~G. and {Sharp}, R.~D.},
  journal  = {Journal of Geophysical Research},
  title    = {{Satellite observations of energetic heavy ions during a geomagnetic storm}},
  year     = {1972},
  month    = jan,
  number   = {31},
  pages    = {6104},
  volume   = {77},
  doi      = {10.1029/JA077i031p06104},
  url      = {https://ui.adsabs.harvard.edu/abs/1972JGR....77.6104S/abstract},
}

@Article{Yau07,
  author   = {{Yau}, Andrew W. and {Abe}, Takumi and {Peterson}, W.K.},
  journal  = {Journal of Atmospheric and Solar-Terrestrial Physics},
  title    = {{The polar wind: Recent observations}},
  year     = {2007},
  month    = nov,
  number   = {16},
  pages    = {1936-1983},
  volume   = {69},
  doi      = {10.1016/j.jastp.2007.08.010},
  url      = {https://ui.adsabs.harvard.edu/abs/2007JASTP..69.1936Y/abstract},
}

@ARTICLE{Terada17,
       author = {{Terada}, Kentaro and {Yokota}, Shoichiro and {Saito}, Yoshifumi and {Kitamura}, Naritoshi and {Asamura}, Kazushi and {Nishino}, Masaki N.},
        title = "{Biogenic oxygen from Earth transported to the Moon by a wind of magnetospheric ions}",
      journal = {Nature Astronomy},
         year = 2017,
        month = jan,
       volume = {1},
          eid = {0026},
        pages = {0026},
          doi = {10.1038/s41550-016-0026},
       adsurl = {https://ui.adsabs.harvard.edu/abs/2017NatAs...1E..26T}
}

@ARTICLE{Hoffman74,
       author = {{Hoffman}, J.~H. and {Dodson}, W.~H. and {Lippincott}, C.~R. and {Hammack}, H.~D.},
        title = "{Initial ion composition results from the Isis 2 satellite}",
      journal = {Journal of Geophysical Research},
         year = 1974,
        month = oct,
       volume = {79},
       number = {28},
        pages = {4246},
          doi = {10.1029/JA079i028p04246},
       adsurl = {https://ui.adsabs.harvard.edu/abs/1974JGR....79.4246H}
}

@Article{Adam00,
  author   = {{Adam}, W. and {Berdermann}, E. and {Bergonzo}, P. and {Bertuccio}, G. and {Bogani}, F. and {Borchi}, E. and {Brambilla}, A. and {Bruzzi}, M. and {Colledani}, C. and {Conway}, J. and {Dangelo}, P. and {Dabrowski}, W. and {Delpierre}, P. and {Deneuville}, A. and {Dulinski}, W. and {van Eijk}, B. and {Fallou}, A. and {Fizzotti}, F. and {Foulon}, F. and {Friedl}, M. and {Gan}, K.~K. and {Gheeraert}, E. and {Grigoriev}, E. and {Hallewell}, G. and {Han}, S. and {Hartjes}, F. and {Hrubec}, J. and {Husson}, D. and {Kagan}, H. and {Kania}, D. and {Kaplon}, J. and {Karl}, C. and {Kass}, R. and {Krammer}, M. and {Logiudice}, A. and {Lu}, R. and {Manfredotti}, C. and {Meier}, D. and {Mishina}, M. and {Moroni}, L. and {Oh}, A. and {Pan}, L.~S. and {Pernicka}, M. and {Peitz}, A. and {Pirollo}, S. and {Polesello}, P. and {Procario}, M. and {Riester}, J.~L. and {Roe}, S. and {Rousseau}, L. and {Rudge}, A. and {Russ}, J. and {Sala}, S. and {Sampietro}, M. and {Schnetzer}, S. and {Sciortino}, S. and {Stelzer}, H. and {Stone}, R. and {Suter}, B. and {Tapper}, R.~J. and {Tesarek}, R. and {Trawick}, M. and {Trischuk}, W. and {Tromson}, D. and {Vittone}, E. and {Walsh}, A.~M. and {Wedenig}, R. and {Weilhammer}, P. and {White}, C. and {Zeuner}, W. and {Zoeller}, M. and {Fenyvesi}, A. and {Molnar}, J. and {Sohler}, D. and {RD42 Collaboration}},
  journal  = {Nuclear Instruments and Methods in Physics Research A},
  title    = {{Pulse height distribution and radiation tolerance of CVD diamond detectors}},
  year     = {2000},
  month    = jun,
  number   = {1-2},
  pages    = {244-250},
  volume   = {447},
  adsurl   = {https://ui.adsabs.harvard.edu/abs/2000NIMPA.447..244A},
  doi      = {10.1016/S0168-9002(00)00195-9},
  url      = {https://ui.adsabs.harvard.edu/abs/2000NIMPA.447..244A/abstract},
}

@Article{Hochedez01,
  author   = {{Hochedez}, J.-F. and {Bergonzo}, P. and {Castex}, M.-C. and {Dhez}, P. and {Hainaut}, O. and {Sacchi}, M. and {Alvarez}, J. and {Boyer}, H. and {Deneuville}, A. and {Gibart}, P. and {Guizard}, B. and {Kleider}, J.-P. and {Lemaire}, P. and {Mer}, C. and {Monroy}, E. and {Mu{\~n}oz}, E. and {Muret}, P. and {Omnes}, F. and {Pau}, J.~L. and {Ralchenko}, V. and {Tromson}, D. and {Verwichte}, E. and {Vial}, J.-C.},
  journal  = {Diamond and Related Materials},
  title    = {{Diamond UV detectors for future solar physics missions}},
  year     = {2001},
  month    = jan,
  number   = {3},
  pages    = {673-680},
  volume   = {10},
  adsurl   = {https://ui.adsabs.harvard.edu/abs/2001DRM....10..673H},
  doi      = {10.1016/S0925-9635(01)00374-0},
  url      = {https://ui.adsabs.harvard.edu/abs/2001DRM....10..673H/abstract},
}

@Article{Hochedez06,
  author   = {{Hochedez}, J.-F. and {Schmutz}, W. and {Stockman}, Y. and {Sch{\"u}hle}, U. and {Benmoussa}, A. and {Koller}, S. and {Haenen}, K. and {Berghmans}, D. and {Defise}, J.-M. and {Halain}, J.-P. and {Theissen}, A. and {Delouille}, V. and {Slemzin}, V. and {Gillotay}, D. and {Fussen}, D. and {Dominique}, M. and {Vanhellemont}, F. and {McMullin}, D. and {Kretzschmar}, M. and {Mitrofanov}, A. and {Nicula}, B. and {Wauters}, L. and {Roth}, H. and {Rozanov}, E. and {R{\"u}edi}, I. and {Wehrli}, C. and {Soltani}, A. and {Amano}, H. and {van der Linden}, R. and {Zhukov}, A. and {Clette}, F. and {Koizumi}, S. and {Mortet}, V. and {Remes}, Z. and {Petersen}, R. and {Nesl{\'a}dek}, M. and {D'Olieslaeger}, M. and {Roggen}, J. and {Rochus}, P.},
  journal  = {Advances in Space Research},
  title    = {{LYRA, a solar UV radiometer on Proba2}},
  year     = {2006},
  month    = jan,
  number   = {2},
  pages    = {303-312},
  volume   = {37},
  adsurl   = {https://ui.adsabs.harvard.edu/abs/2006AdSpR..37..303H},
  doi      = {10.1016/j.asr.2005.10.041},
  url      = {https://ui.adsabs.harvard.edu/abs/2006AdSpR..37..303H/abstract},
}

@Article{Hochedez00,
  author  = {{Hochedez}, J.-F. and {Verwichte}, E. and {Bergonzo}, P. and {Guizard}, B. and {Mer}, C. and {Tromson}, D. and {Sacchi}, M. and {Dhez}, P. and {Hainaut}, O. and {Lemaire}, P. and {Vial}, J.-C.},
  journal = {Physica Status Solidi Applied Research},
  title   = {{Future Diamond UV Imagers For Solar Physics}},
  year    = {2000},
  month   = jan,
  number  = {1},
  pages   = {141-149},
  volume  = {181},
  adsurl  = {https://ui.adsabs.harvard.edu/abs/2000PSSAR.181..141H},
  doi     = {10.1002/1521-396X(200009)181:1<141::AID-PSSA141>3.0.CO;2-B},
  url     = {https://ui.adsabs.harvard.edu/abs/2000PSSAR.181..141H/abstract},
}

@Article{Benmoussa03,
  author   = {{Benmoussa}, A. and {Hochedez}, J.-F. and {Schmutz}, W.~K. and {Sch{\"u}hle}, U. and {Nesl{\'a}dek}, M. and {Stockman}, Y. and {Kroth}, U. and {Richter}, M. and {Theissen}, A. and {Remes}, Z. and {Haenen}, K. and {Mortet}, V. and {Koller}, S. and {Halain}, J.~P. and {Petersen}, R. and {Dominique}, M. and {D'Olieslaeger}, M.},
  journal  = {Experimental Astronomy},
  title    = {{Solar-Blind Diamond Detectors for Lyra, the Solar VUV Radiometer on Board Proba II}},
  year     = {2003},
  month    = dec,
  number   = {3},
  pages    = {141-148},
  volume   = {16},
  adsurl   = {https://ui.adsabs.harvard.edu/abs/2003ExA....16..141B},
  doi      = {10.1007/s10686-003-0035-3},
  url      = {https://ui.adsabs.harvard.edu/abs/2003ExA....16..141B/abstract},
}

@Article{BenMoussa06,
  author   = {{BenMoussa}, A. and {Hochedez}, J.~F. and {Sch{\"u}hle}, U. and {Schmutz}, W. and {Haenen}, K. and {Stockman}, Y. and {Soltani}, A. and {Scholze}, F. and {Kroth}, U. and {Mortet}, V. and {Theissen}, A. and {Laubis}, C. and {Richter}, M. and {Koller}, S. and {Defise}, J.-M. and {Koizumi}, S.},
  journal  = {Diamond and Related Materials},
  title    = {{Diamond detectors for LYRA, the solar VUV radiometer on board PROBA2}},
  year     = {2006},
  month    = jan,
  number   = {4},
  pages    = {802-806},
  volume   = {15},
  adsurl   = {https://ui.adsabs.harvard.edu/abs/2006DRM....15..802B},
  doi      = {10.1016/j.diamond.2005.10.024},
  url      = {https://ui.adsabs.harvard.edu/abs/2006DRM....15..802B/abstract},
}

@Article{Pace03,
  author   = {{Pace}, E. and {De Sio}, A.},
  journal  = {Nuclear Instruments and Methods in Physics Research A},
  title    = {{Diamond detectors for space applications}},
  year     = {2003},
  month    = nov,
  number   = {1-3},
  pages    = {93-99},
  volume   = {514},
  adsurl   = {https://ui.adsabs.harvard.edu/abs/2003NIMPA.514...93P},
  doi      = {10.1016/j.nima.2003.08.088},
  url      = {https://ui.adsabs.harvard.edu/abs/2003NIMPA.514...93P/abstract},
}

@Article{Mendoza15,
  author   = {{Mendoza}, Frank and {Makarov}, Vladimir and {Weiner}, Brad R. and {Morell}, Gerardo},
  journal  = {Applied Physics Letters},
  title    = {{Solar-blind field-emission diamond ultraviolet detector}},
  year     = {2015},
  month    = nov,
  number   = {20},
  pages    = {201605},
  volume   = {107},
  adsurl   = {https://ui.adsabs.harvard.edu/abs/2015ApPhL.107t1605M},
  doi      = {10.1063/1.4936162},
  eid      = {201605},
  url      = {https://ui.adsabs.harvard.edu/abs/2015ApPhL.107t1605M/abstract},
}

@Article{DeSio03,
  author   = {{De Sio}, A. and {Donato}, M.~G. and {Faggio}, G. and {Marinelli}, Marco and {Messina}, G. and {Milani}, E. and {Pace}, E. and {Paoletti}, A. and {Pini}, A. and {Santangelo}, S. and {Scuderi}, S. and {Tucciarone}, A. and {Verona-Rinati}, G.},
  journal  = {Diamond and Related Materials},
  title    = {{Spectral response of large area CVD diamond photoconductors for space applications in the vacuum UV}},
  year     = {2003},
  month    = jan,
  number   = {10},
  pages    = {1819-1824},
  volume   = {12},
  adsurl   = {https://ui.adsabs.harvard.edu/abs/2003DRM....12.1819D},
  doi      = {10.1016/S0925-9635(03)00216-4},
  url      = {https://ui.adsabs.harvard.edu/abs/2003DRM....12.1819D/abstract},
}

@INPROCEEDINGS{Heynderickx03, author={Heynderickx, D. and Quaghebeur, B. and Wera, J. and Daly, E.J. and Evans, H.D.R.}, booktitle={Proceedings of the 7th European Conference on Radiation and Its Effects on Components and Systems, 2003. RADECS 2003.}, title={New radiation environment and effects models in ESA's space environment information system (SPENVIS)}, year={2003}, volume={}, number={}, pages={643-646}, doi={}}

@inbook{Heynderickx12,
author = {D. Heynderickx and B. Quaghebeur and E. Speelman and E. Daly},
title = {ESA's Space Environment Information System (SPENVIS) - A WWW interface to models of the space environment and its effects},
booktitle = {38th Aerospace Sciences Meeting and Exhibit},
chapter = {},
pages = {},
doi = {10.2514/6.2000-371},
URL = {https://arc.aiaa.org/doi/abs/10.2514/6.2000-371},
eprint = {https://arc.aiaa.org/doi/pdf/10.2514/6.2000-371}
}

@Misc{Berger92,
  author   = {{Berger}, M.~J.},
  month    = dec,
  title    = {{ESTAR, PSTAR, and ASTAR: Computer programs for calculating stopping-power and range tables for electrons, protons, and helium ions}},
  year     = {1992},
  adsurl   = {https://ui.adsabs.harvard.edu/abs/1992esta.rept.....B},
  url      = {https://ui.adsabs.harvard.edu/abs/1992esta.rept.....B/abstract},
}

@InProceedings{Kruglanski09,
  author    = {{Kruglanski}, M. and {Messios}, N. and {de Donder}, E. and {Gamby}, E. and {Calders}, S. and {Hetey}, L. and {Evans}, H.},
  booktitle = {EGU General Assembly Conference Abstracts},
  title     = {{Space Environment Information System (SPENVIS)}},
  year      = {2009},
  month     = apr,
  pages     = {7457},
  series    = {EGU General Assembly Conference Abstracts},
  adsurl    = {https://ui.adsabs.harvard.edu/abs/2009EGUGA..11.7457K},
  url       = {https://ui.adsabs.harvard.edu/abs/2009EGUGA..11.7457K/abstract},
}

@Book{Janesick01,
  author    = {{Janesick}, J.~R.},
  publisher = {{Society of Photo Optical}},
  title     = {{Scientific charge-coupled devices}},
  year      = {2001},
  adsurl    = {http://adsabs.harvard.edu/abs/2001sccd.book.....J},
  booktitle = {Scientific charge-coupled devices, Bellingham, WA: SPIE Optical Engineering Press, 2001, xvi, 906 p.~SPIE Press monograph, PM 83.~ISBN 0819436984},
  url       = {http://adsabs.harvard.edu/abs/2001sccd.book.....J},
}

@ARTICLE{Ueda13,
  author = {{Ueda}, S. and {Hayashida}, K. and {Nakajima}, H. and {Anabuki},
	N. and {Tsunemi}, H. and {Kan}, H. and {Kohmura}, T. and {Ikeda},
	S. and {Kaneko}, K. and {Watanabe}, T. and {Mori}, K. and {Nobukawa},
	M. and {Murakami}, H. and {Sakata}, K. and {Todoroki}, S. and {Yagihashi},
	N. and {Mizuno}, E. and {Muramatsu}, M. and {Suzuki}, H. and {Takagi},
	S.},
  title = {{Measurement of the soft X-ray response of P-channel back-illuminated
	CCD}},
  journal = {Nuclear Instruments and Methods in Physics Research A},
  year = {2013},
  volume = {704},
  pages = {140-146},
  month = {mar},
  adsurl = {http://adsabs.harvard.edu/abs/2013NIMPA.704..140U},
  doi = {10.1016/j.nima.2012.11.187},
  url = {https://ui.adsabs.harvard.edu/abs/2013NIMPA.704..140U/abstract}
}

@ARTICLE{Inoue16,
  author = {{Inoue}, S. and {Hayashida}, K. and {Katada}, S. and {Nakajima},
	H. and {Nagino}, R. and {Anabuki}, N. and {Tsunemi}, H. and {Tsuru},
	T.~G. and {Tanaka}, T. and {Uchida}, H. and {Nobukawa}, M. and {Nobukawa},
	K.~K. and {Washino}, R. and {Mori}, K. and {Isoda}, E. and {Sakata},
	M. and {Kohmura}, T. and {Tamasawa}, K. and {Tanno}, S. and {Yoshino},
	Y. and {Konno}, T. and {Ueda}, S.},
  title = {{Modeling the spectral response for the soft X-ray imager onboard
	the ASTRO-H satellite}},
  journal = {Nuclear Instruments and Methods in Physics Research A},
  year = {2016},
  volume = {831},
  pages = {415-419},
  month = sep,
  adsurl = {http://adsabs.harvard.edu/abs/2016NIMPA.831..415I},
  doi = {10.1016/j.nima.2016.03.071},
  url = {http://adsabs.harvard.edu/abs/2016NIMPA.831..415I}
}

@Article{Ichikawa24,
  author   = {{Ichikawa}, Kimiyoshi and {Matsumoto}, Tsubasa and {Inokuma}, Takao and {Yamasaki}, Satoshi and {Nebel}, Christoph E. and {Tokuda}, Norio},
  journal  = {Accounts of Materials Research},
  title    = {{Diamond Homoepitaxial Growth Technology toward Wafer Fabrication, Atomically Controlled Surfaces, and Low Resistivity}},
  year     = {2024},
  month    = oct,
  number   = {10},
  pages    = {1181-1193},
  volume   = {5},
  adsurl   = {https://ui.adsabs.harvard.edu/abs/2024AMatR...5.1181I},
  doi      = {10.1021/accountsmr.4c00123},
  url      = {https://ui.adsabs.harvard.edu/abs/2024AMatR...5.1181I/abstract},
}

@Article{Pomorski06,
  author   = {{Pomorski}, M. and {Berdermann}, E. and {Caragheorgheopol}, A. and {Ciobanu}, M. and {Ki}, M. and {Martemiyanov}, A. and {Nebel}, C. and {Moritz}, P.},
  journal  = {Physica Status Solidi Applied Research},
  title    = {{Development of single-crystal CVD-diamond detectors for spectroscopy and timing}},
  year     = {2006},
  month    = sep,
  number   = {12},
  pages    = {3152-3160},
  volume   = {203},
  doi      = {10.1002/pssa.200671127},
  url      = {https://ui.adsabs.harvard.edu/abs/2006PSSAR.203.3152P/abstract},
}

@Article{Shimaoka16,
  author   = {{Shimaoka}, Takehiro and {Kaneko}, Junichi H. and {Tsubota}, Masakatsu and {Shimmyo}, Hiroaki and {Watanabe}, Hideyuki and {Chayahara}, Akiyoshi and {Umezawa}, Hitoshi and {Shikata}, Shin-ichi},
  journal  = {EPL (Europhysics Letters)},
  title    = {{High-performance diamond radiation detectors produced by lift-off method}},
  year     = {2016},
  month    = mar,
  number   = {6},
  pages    = {62001},
  volume   = {113},
  adsurl   = {https://ui.adsabs.harvard.edu/abs/2016EL....11362001S},
  doi      = {10.1209/0295-5075/113/62001},
  url      = {https://ui.adsabs.harvard.edu/abs/2016EL....11362001S/abstract},
}

@Article{Shimaoka21,
  author    = {T. Shimaoka and S. Koizumi and J. H. and Kaneko},
  journal   = {Functional Diamond},
  title     = {Recent progress in diamond radiation detectors},
  year      = {2021},
  number    = {1},
  pages     = {205--220},
  volume    = {1},
  doi       = {10.1080/26941112.2021.2017758},
  eprint    = {https://doi.org/10.1080/26941112.2021.2017758},
  publisher = {Taylor \& Francis},
  url       = {https://doi.org/10.1080/26941112.2021.2017758

    https://www.tandfonline.com/doi/full/10.1080/26941112.2021.2017758#abstract},
}

@Article{Gabrysch11,
  author   = {{Gabrysch}, Markus and {Majdi}, Saman and {Twitchen}, Daniel J. and {Isberg}, Jan},
  journal  = {Journal of Applied Physics},
  title    = {{Electron and hole drift velocity in chemical vapor deposition diamond}},
  year     = {2011},
  month    = mar,
  number   = {6},
  pages    = {063719-063719-4},
  volume   = {109},
  adsurl   = {https://ui.adsabs.harvard.edu/abs/2011JAP...109f3719G},
  doi      = {10.1063/1.3554721},
  eid      = {063719-063719-4},
  url      = {https://ui.adsabs.harvard.edu/abs/2011JAP...109f3719G/abstract},
}

@Article{Kholili24,
  author   = {{Kholili}, Muhammad Jauhar and {Shimaoka}, Takehiro and {Hara}, Asuka and {Tanaka}, Manobu M.},
  journal  = {Physica Status Solidi Applied Research},
  title    = {{Transient-Current Technique Characterization of Diamond Detector and Proposal of a Diamond Detector with Charge Amplification}},
  year     = {2024},
  month    = feb,
  number   = {4},
  pages    = {2300673},
  volume   = {221},
  adsurl   = {https://ui.adsabs.harvard.edu/abs/2024PSSAR.22100673K},
  doi      = {10.1002/pssa.202300673},
  eid      = {2300673},
  url      = {https://ui.adsabs.harvard.edu/abs/2024PSSAR.22100673K/abstract},
}

@Article{Rahman26,
  author        = {{Rahman Ishaqzai}, Faiz and {Deniz}, Muhammed and {Kr{\"o}ninger}, Kevin and {Weingarten}, Jens},
  journal       = {arXiv e-prints},
  title         = {{Charge-Carrier Mobility in Diamond: Review, Data Compilation, and Modelling for Detector Simulations}},
  year          = {2026},
  month         = jan,
  pages         = {arXiv:2601.10807},
  adsurl        = {https://ui.adsabs.harvard.edu/abs/2026arXiv260110807R},
  archiveprefix = {arXiv},
  doi           = {10.48550/arXiv.2601.10807},
  eid           = {arXiv:2601.10807},
  eprint        = {2601.10807},
  primaryclass  = {physics.ins-det},
  url           = {https://ui.adsabs.harvard.edu/abs/2026arXiv260110807R/abstract},
}

\bio{}
\endbio


\end{document}